\documentclass{ws-procs9x6}

\begin{document}

\title{DEEPLY VIRTUAL EXCLUSIVE PROCESSES WITH CHARM}

\author{S. LIUTI$^1$}

\address{%
University of Virginia - Physics Department \\
382, McCormick Rd., Charlottesville, Virginia 22904 - USA \\ 
$^1$ E-mail: \email{sl4y@virginia.edu}} 

\author{G.R. GOLDSTEIN$^2$}

\address{%
Department of Physics and Astronomy \\
Tufts University, Medford, MA 02155 - USA. \\
$^2$ E-mail: \email{gary.goldstein@tufts.edu}}

\begin{abstract}
We propose to investigate a largely unexplored sector that is unique to the formulation 
of hard exclusive processes in terms of GPDs, namely the electroproduction of strange and charmed mesons in the
kinematical ranges of Jefferson Lab's 12 GeV upgrade, 
and of the proposed Electron Ion Collider (EIC). In this contribution we focus on charmed meson production that is unique to the EIC.
Exclusive strange and charmed meson production provides new insights in the 
connection of the quark/gluon 
degrees of freedom with the meson-baryon description, both in the unpolarized and polarized sectors. 
However, as particularly evident in polarized scattering, the underlying mechanisms are still far from being fully understood. 
We present an approach in terms of generalized parton distributions. As an application, we show that  through exclusive electroproduction of pseudoscalar charmed mesons one can uniquely single out the non-perturbative charmed component in the nucleon structure function.
\end{abstract}


\bodymatter

\section{Introduction}\label{sec1}
Many experiments have been conducted to determine the charm content of the proton.
Different types of hadronic reactions allow us in principle to access the charm quark distributions in a varied set of kinematical ranges, from hadron scattering at the Tevatron to inclusive and semi-inclusive leptoproduction at colliders (HERA) and in various fixed target experiments.

\begin{figure}
\centering
\centerline{\includegraphics[width=0.50\textwidth]{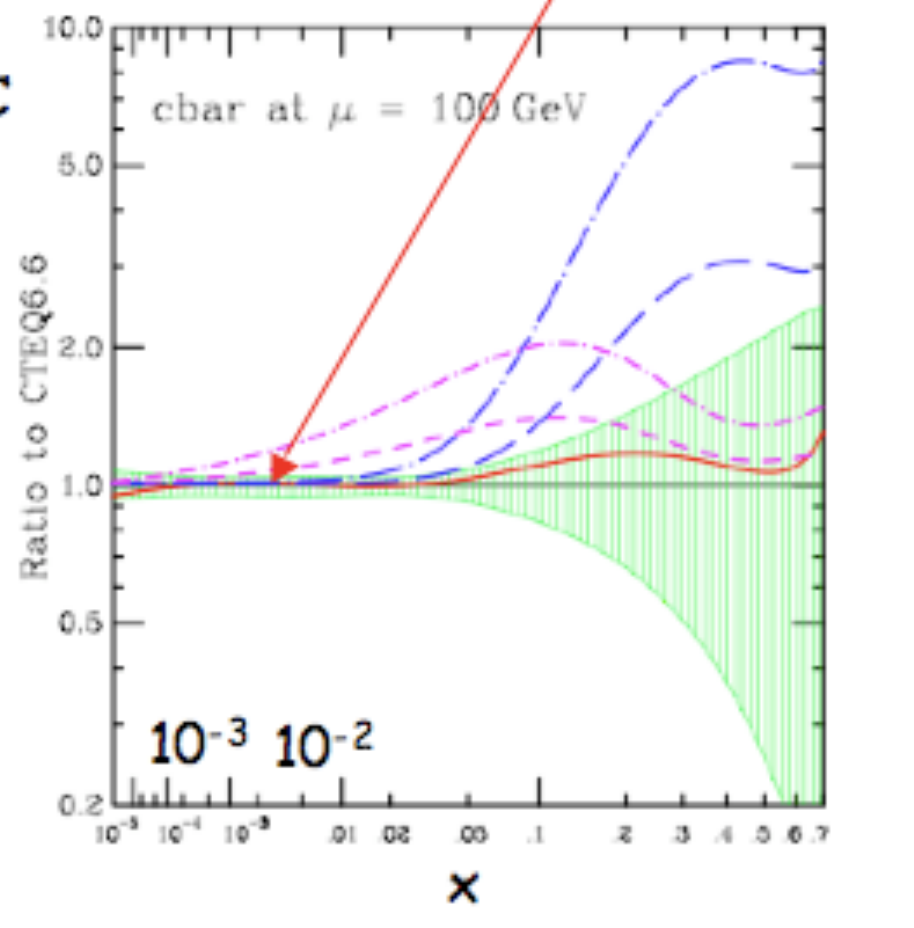}}
\caption{Illustration of the status of present studies of the charmed quarks parton distribution.  The arrow, and the visibly smaller error band, indicate the region in Bjorken $x$ were most experiments are concentrated. The dashed and dot-dashed curves were obtained including different models of IC at the scale $\mu=100$ GeV  (adapted from \refcite{Nad}). }
	\label{fig0}
\end{figure}

Two main competing interpretations can be given on the way charm gets excited. Either $c- \bar{c}$ pairs  are produced through a Perturbative QCD (PQCD) mechanism such as gluon-gluon or photon-gluon fusion, when the scale of 
the probe, $Q^2 \geq 4 m_c^2$, $m_c$ being the charm quark mass, or such pairs exist in the nucleon as a so-called "Intrinsic Charm" (IC) component, with a non-negligible probability density at $Q^2= m_c^2$, due to some underlying non-perturbative mechanism. Predictions from the  two mechanisms at $Q^2 > m_c^2$, could differ in an observable
way.  
The problem of disentangling them lies, however, unsolved to date. Figure \ref{fig0} illustrates the rather large impact of non-perturbative charm  contributions on the state of the art PDF parametrizations (dot-dashed curves) compared to the radiatively generated ones (shaded area) (Ref.\refcite{Nad}).

It has now become particularly pressing to study the heavy quark components of the nucleon because of the advent of the LHC.  The heavy quark components will be key in the study
of QCD matrix elements in the unprecedented multi-TeV CM energy regimes accessible at the LHC.
At the same time, as the LHC opens new horizons for studies of physics beyond the Standard Model,  many ``candidate theories'' will provide similar signatures of a departure from SM predictions.
For these types of precision measurements it will be necessary to provide accurately determined QCD inputs.

The analyses in Ref.\refcite{Pum} have shown how the inclusion of IC could modify the outcome of  global PDF analyses. However, the situation is not clear-cut. PDF analyses do not provide direct evidence that IC exists, but can be considered as indications in that direction. 

\begin{figure}
\centering
\centerline{\includegraphics[width=4.5cm]{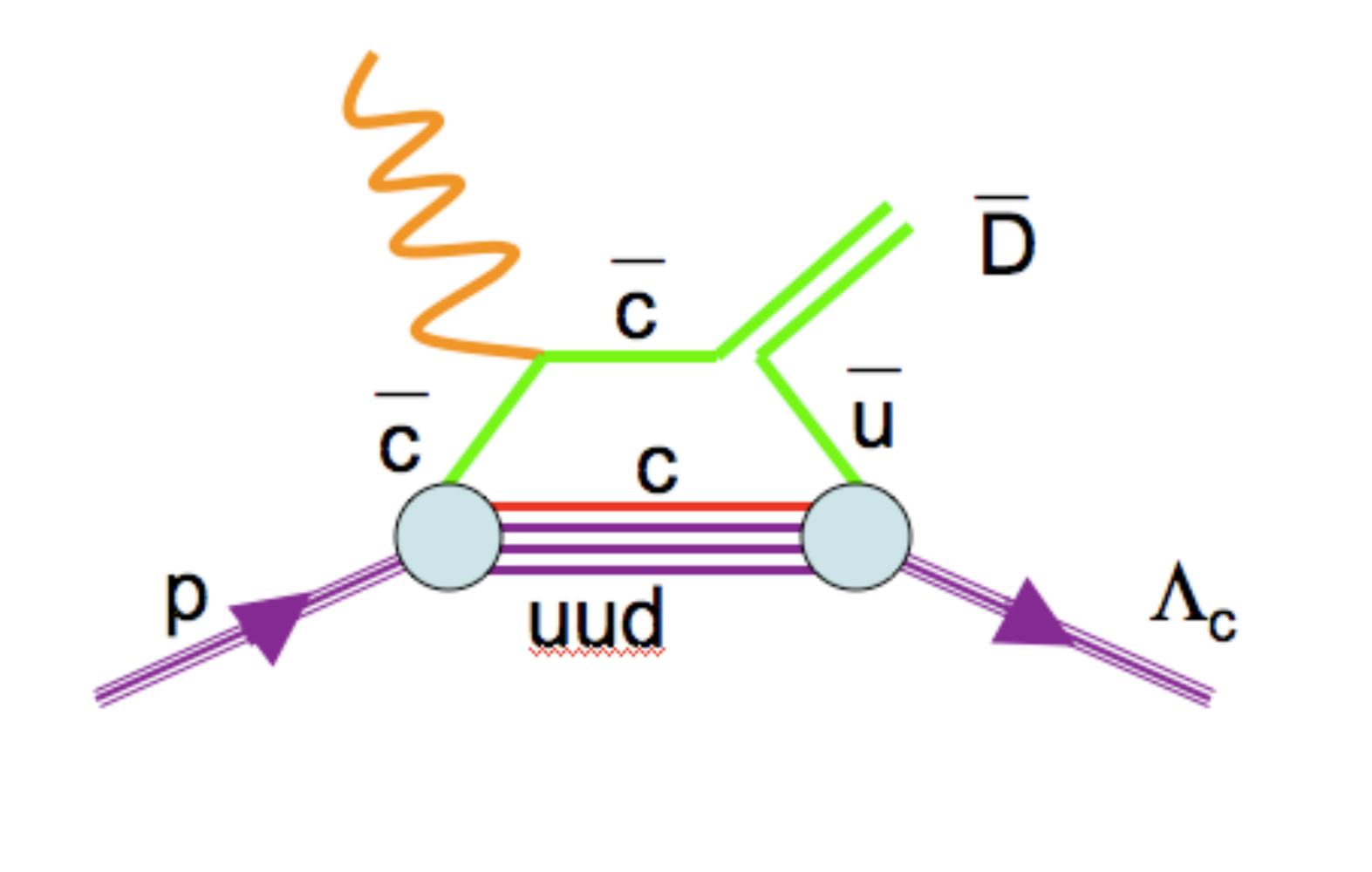}
\includegraphics[width=4.5cm]{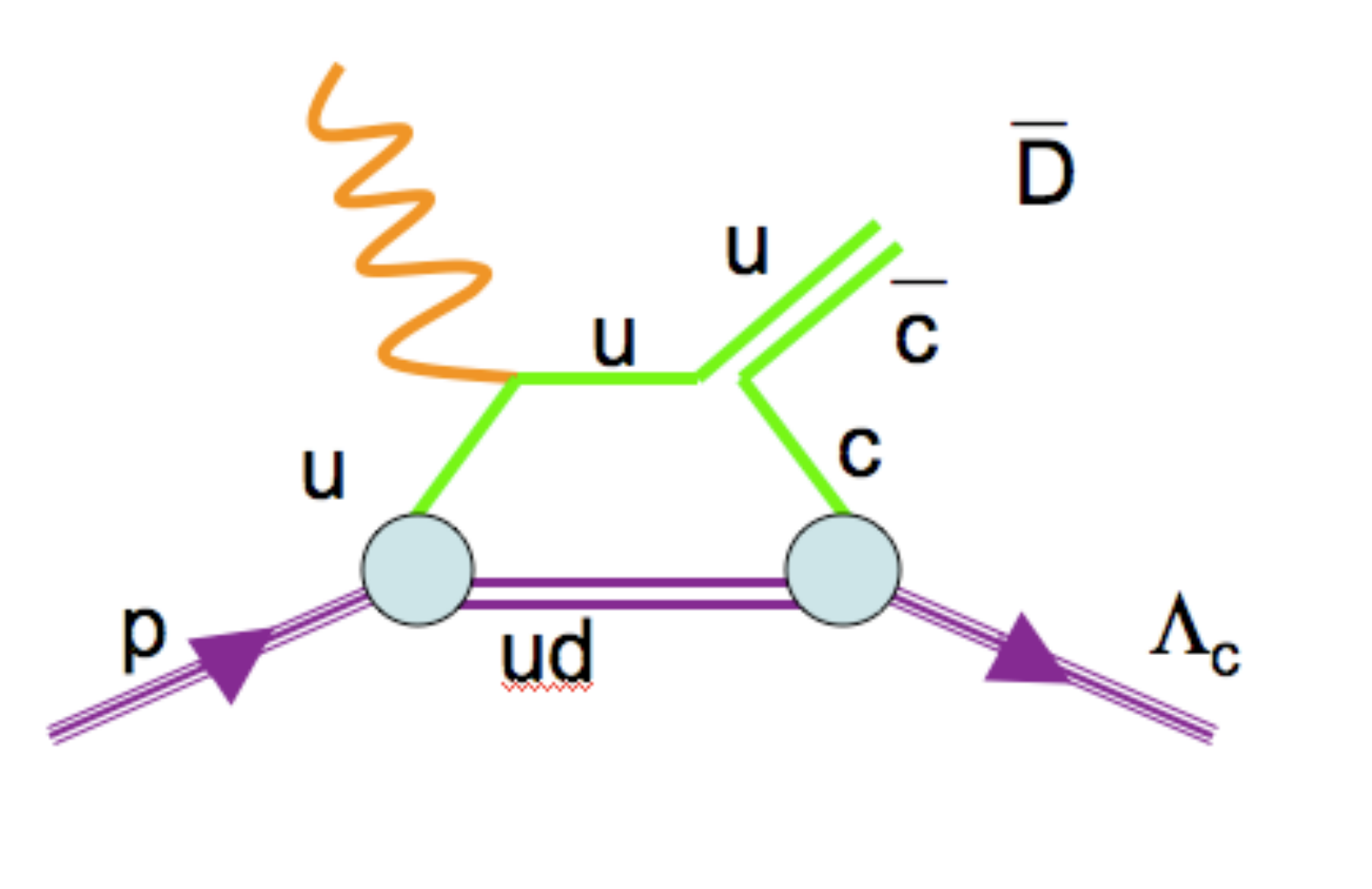}}
\centerline{\includegraphics[width=4.5cm]{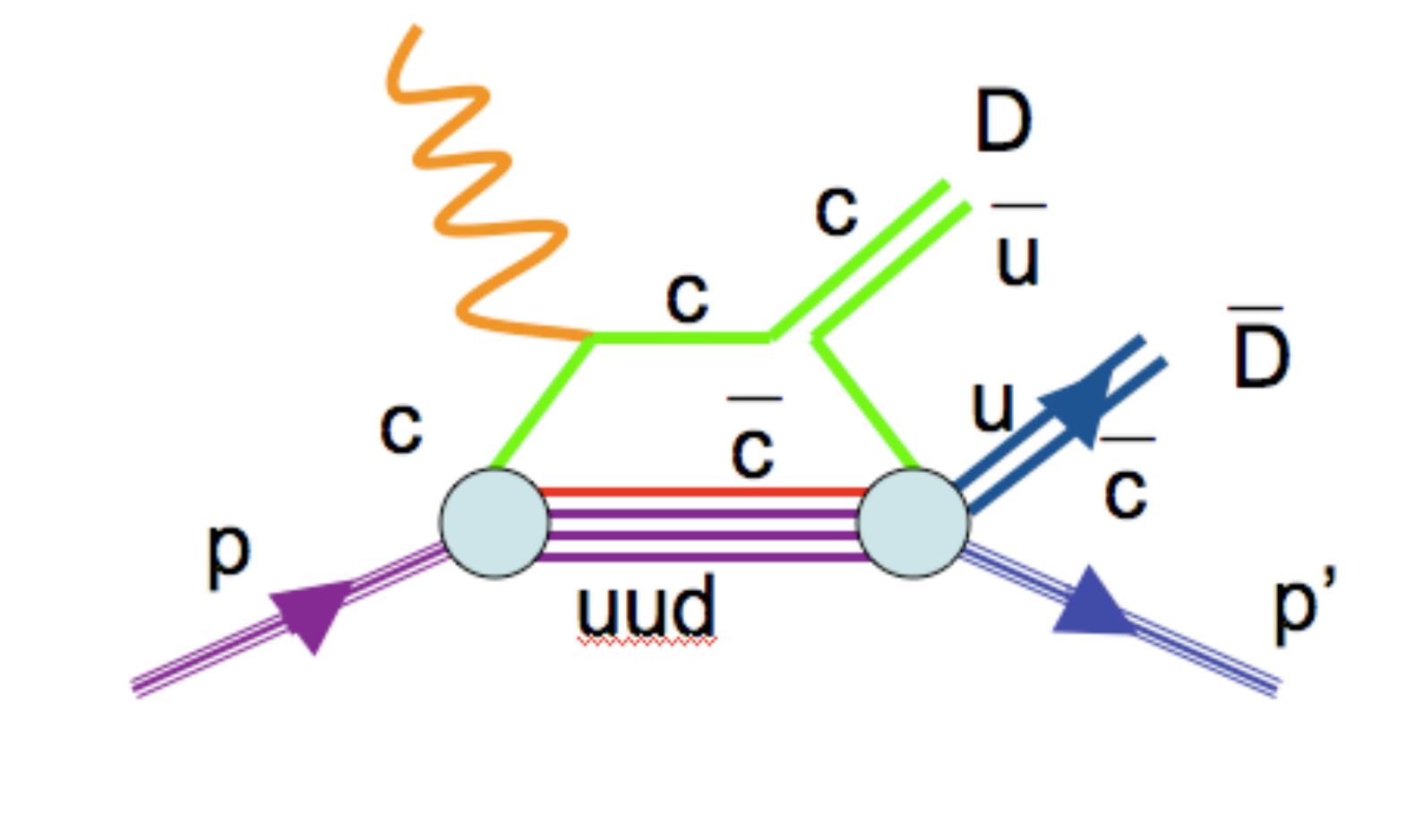}
\includegraphics[width=4.5cm]{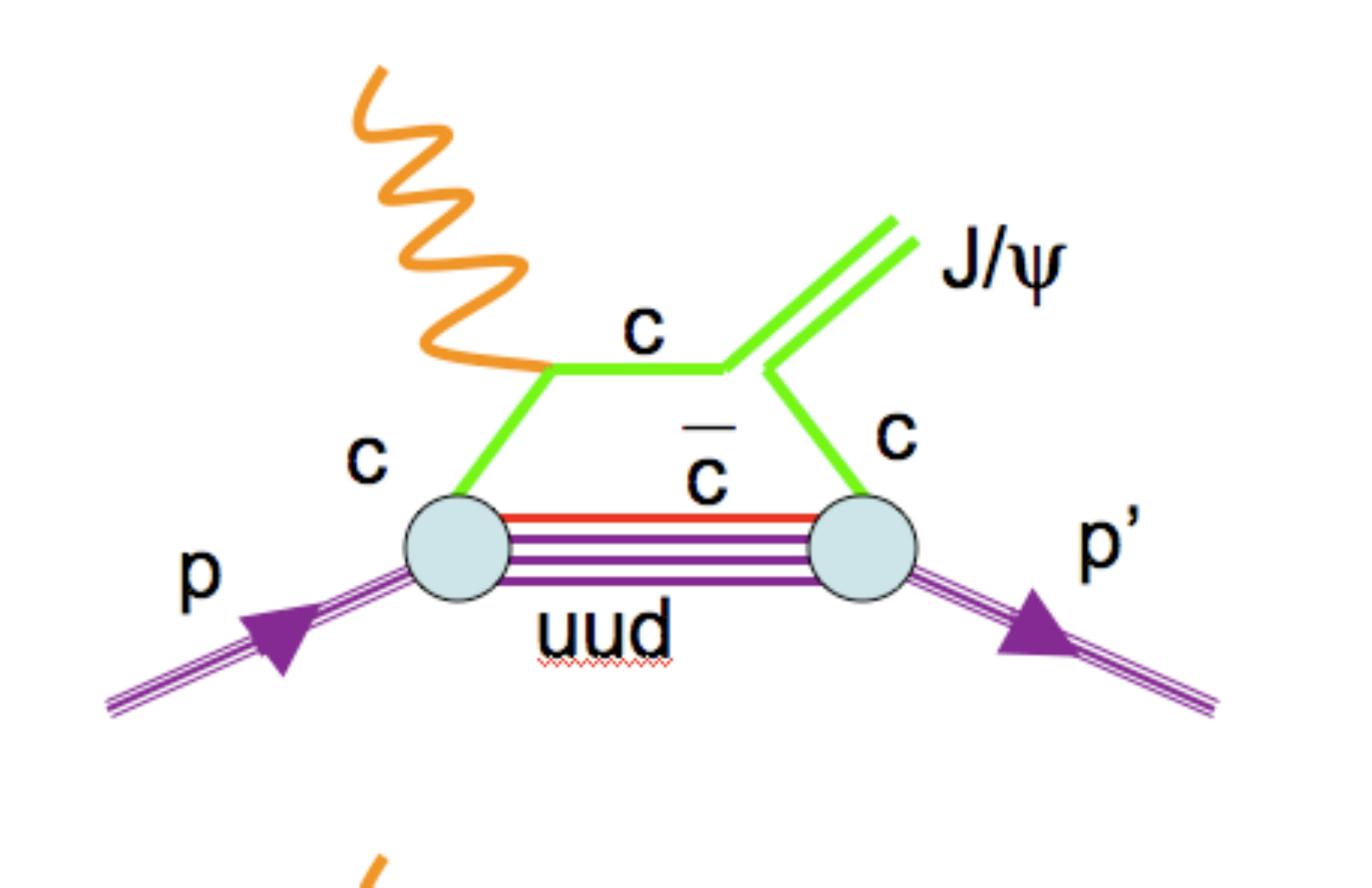}}
\caption{Deeply Virtual Charmed Meson Production (DVCMP). All produced mesons are 
unique probes of the IC charm content.}
\label{fig1}
\end{figure}

In this contribution we make the point that in order to substantiate these findings it is important to identify new observables that would allow us to extend the studies to parton distribution size observables such as in Deeply Virtual Meson Production (DVMP),  and to spin correlations. 
We present preliminary results involving the following electroproduction exclusive processes (Fig.\ref{fig1}): 

\vspace{0.3cm}
\noindent 
(1) $\gamma^* p \rightarrow J/\psi \, p^\prime$, 

\noindent 
(2) $\gamma^* p \rightarrow D \, \overline{D} \, p^\prime$,  

\noindent
(3) $\gamma^* p \rightarrow \overline{D} \, \Lambda_c$,

\noindent
(4) $\gamma^* p \rightarrow \eta_C \, p^\prime$.
\vspace{0.3cm}

As we explain on what follows, these processes necessitate: {\it i)}  high luminosity because they are exclusive; {\it ii)}  high enough $Q^2$, in order to produce the various charmed mesons, and {\it iii)} a wide kinematical range in Bjorken $x$ (see discussion relative to Fig.\ref{fig0}). Their ideal realization would be at an Electron Ion Collider (EIC) that could guarantee high luminosity conditions also at relatively large $x$.

\section{Windows into Heavy Flavor Production at the EIC}
\label{sec2} 
QCD factorization for exclusive meson electroproduction was proven initially in Ref.\refcite{Collins}.  Factorization arguments can be extended to the production of charmed mesons, thus allowing us  to write the various contributions to the cross section, originating from the amplitudes for different beam and target spin configurations, as a convolution over the longitudinal momentum fraction $X$ of the hard scattering process with the quark-proton amplitudes.  
Using the approach in Ref.\refcite{AGL} (see Section \ref{sec3}), we can describe the exclusive production of pseudoscalar charmed mesons mainly in terms of {\em chiral-odd} Generalized Parton Distributions (GPDs) for the $u,d$ and $c$ quarks, $F_T^q \equiv \{H_T\, E_T, \widetilde{H}_T, \widetilde{E}_T\}$, with $q=u,d,c$. 
\footnote{A small contribution from the chiral even GPD is in principle also present, that we disregard here.} 
These by definition do not couple to gluons, or they evolve perturbatively similarly to the flavor non singlet distributions, Refs.\refcite{Bar,Vog,Kum}.
All GPDs involved in pseudoscalar charmed mesons electroproduction are therefore of the "intrinsic" kind since they cannot evolve from gluons. By finding observables that can isolate the charmed GPDs, one can therefore uniquely gauge the size of the IC contribution.    

A somewhat similar perspective was presented in Ref.\refcite{Ana} where it was shown that the IC content of the proton can be studied by measuring asymmetries in inclusive heavy quark jet production, 
$e p \rightarrow e^\prime Q  X(\overline{Q})$. 
In fact, IC does not contribute at all to azimuthal asymmetries of the $cos(2\phi)$ form $\frac{\sigma_{TT}}{\sigma_T + \epsilon \sigma_L}$. Hence the reduction at high $x_{Bj}$ of large asymmetry from their perturbative gluon model signals the intrinsic charm contribution. This, however is dependent on their model and requires measuring an asymmetry that is reduced from the gluon contribution.
While these studies present yet another opportunity for studying the multifaceted and still elusive connections between semi-inclusive and exclusive processes, our approach is a more stringent test for IC since the suggested exclusive production processes occur solely via the IC channel.  

The hard scattering amplitudes for the four processes above are respectively (1)  $\gamma^* c (\bar{c}) \rightarrow  J/\psi \, c (\bar{c})$, (2) $\gamma^* \, c \rightarrow  D \, u$, (3)
$\gamma^* \bar{c}  \rightarrow  \overline{D} \, (\bar{u})$ and $\gamma^* \ u \rightarrow  \overline{D} \, c$, (4) $\gamma^* c (\bar{c}) \rightarrow  \eta_C \, c (\bar{c})$ (Fig.\ref{fig1}). All of these involve the charm mass
scale in some way, along with a heavy meson distribution amplitude. For $D$ and $\bar{D}$ creation in the beam direction, the distribution or wavefunction will be of the same order for each. The $J/\Psi$ will be more difficult to produce in the hard process, aside from its sizable diffractive production, which occurs through the virtual photon dissociation into $c\bar{c}$. The pair scatter diffractively and recombine as the $J/\Psi$.  The diffractive scattering is implemented through two gluon exchange, so that the overall process would involve gluon-nucleon GPDs rather than intrinsic charm, at least in the diffractive region.

Amplitudes (1) and (4) could be smaller than (2) and (3) for several reasons. From the GPD perspective, the soft contribution is given only by the charm GPD, $F_c$ that, based on the constraint from inclusive scattering -- giving the forward limit  of GPDs--  we expect to be a few per cent of the light quarks' GPDs.
Moreover process (4) requires the detection of an $\eta_c$ meson, which seems to be a forbidding task even in a future EIC.
Process (2) does not seem the best way to access GPDs because of the exotic $\overline{D} \, p$ final hadronic configuration.
Finally, process (3) can occur in the two configurations described in Fig.1c and Fig.1d. By assuming $SU(4)$ symmetry, these are proportional to linear combinations of the non-singlet GPDs $H_c, \,H_u$ and $H_d$. So the mechanism involving IC is dominated by scattering off the light flavor quarks, at least before symmetry breaking is implemented. 
However, by detecting different baryons in the final state, that is considering the reactions:

\vspace{0.3cm}
\noindent $\gamma^* p  \rightarrow  \overline{D}^0 \, \Lambda_c^+ \Rightarrow \left(2 F_T^u - F_T^d + F_T^c\right)/\sqrt{6}$, 

\noindent $\gamma^* p  \rightarrow  \overline{D}^0 \, \Sigma_c^+ \Rightarrow \left(F_T^d - F_T^c\right)/\sqrt{2}$, 

\noindent 
$\gamma^* n  \rightarrow  D^- \, \Sigma_c^{++} \Rightarrow F_T^d - F_T^c$,
\vspace{0.3cm}

\noindent
one can extract the charmed GPDs, $F_T^c$.

We related the GPDs that involve charmed quarks or charmed hadrons  to their uncharmed counterparts through flavor SU(4). In the GPDs involving open charm production, process (3), there are the off-diagonal amplitudes for $N\rightarrow u\,(d):c\rightarrow \Lambda_C^+,\,\Sigma_C^+\,(\Sigma_C^{++})$. The spin $\frac{1}{2}$ hadrons belong to the 20-plet. To simplify the determination of the SU(4) Clebsch-Gordon coefficients, notice that the $n\, \& \, p$
are in the same relation to these 3 charmed baryons as the strange set $\Lambda^0,\, \Sigma^0,$ and $\Sigma^+$, which corresponds to replacing c by s and vice versa. 
This result can be seen as a rotation in the SU(4) space of the usual baryon octet with no s, one s, two s (the non-charm subset of the 20) to the non-strange no c, one c and two c octet (the non-strange subset of the 20). 
The same correspondence applies to the meson 15-plet. The 4 X 4 matrix can have s \& c entries reversed. Now there is strong breaking of the SU(4) symmetry. 
There are many ans{\"a}tze for such symmetry breaking, e.g. Ref.~\refcite{Song}. 
Obviously, symmetry breaking must suppress the production of charm. Phase space does enter, but other effects may be significant. This is being studied.

\section{Pseudoscalar Charmed Mesons Electroproduction and Chiral Odd GPDs} 
\label{sec3}
The next step is to provide a model for charmed GPDs that could guide experimenters for future simulations and possible extractions of $F_T^c$ from data. Our approach is similar to Refs.\refcite{AHLT1,AHLT2}, where a Regge improved quark-diquark model was devised to provide a parametrization of the chiral even, unpolarized GPDs $H^q$ and $E^q$, $q=u,d$ that is quantitatively constrained by the nucleon form factors data, and from deep inelastic scattering structure functions data through
\[ \int_{-1+ \zeta}^1 dX H^q(X,\zeta,t) = F_1^q(t) \; \; \; \; \; \int_{-1+ \zeta}^1dX  E^q(X,\zeta,t) = F_2^q(t)\]

\[ F^q(X,0,0; Q^2) \equiv   q(X,Q^2) \] 
where we introduced the kinematical variables $X = k^+/P^+$, $\zeta=\Delta^+/P^+$, and $t= \Delta^2$, $P$ being the proton momentum, $k$ the struck quark momentum, and $\Delta$ the momentum transfer between the initial and final proton. $F_{1(2)}^q$ are the quark contributions to the nucleon Dirac and Pauli form factors, and $q(X)$ is the quark parton distribution at a given scale, $Q^2$ (for a review of GPDs see Ref.\refcite{BelRad}).

\begin{figure}
\centering
\centerline{\includegraphics[width=6.cm]{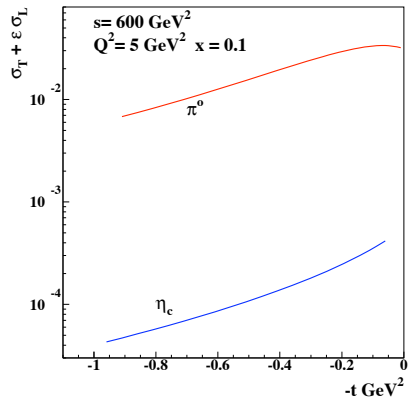}}
\caption{Comparison of $\pi^o$ and $\eta_c$ cross sections. Although $\eta_c$ is hard to detect,
the range between the two lines gives an estimate of where the cross sections for the other processes
will lie.}
\label{fig3}
\end{figure}

In order to constrain charmed GPD, $H_T^{c(\bar{c})}$, we used the same set of non-perturbative models of IC that were used in Ref.\refcite{Pum}. These were found to be consistent with present DIS data, posing an upper limit on the intrinsic charm content measured by the momentum fraction $\langle x \rangle_{c+\bar{c} }
\leq 2 \%$. 
The constraint from the form factor is less obvious since no measurements or {\it ab initio} evaluations are available
(see however, Ref.\refcite{lat}). We therefore took as an upper limit the recent experimental extraction of 
the strangeness axial form factor $G_A^s(t)$ [\refcite{Pate}]. This predicts negative values of $G_A^s$ in the range $0.47 \leq -t \leq 1$ GeV$^2$, that are consistent with recent findings of $\Delta s$. 
Initial results based on these estimates are shown in Fig.\ref{fig3}.
%
We have chiral odd GPDs for these pseudoscalar processes, just as we introduced for the $\pi^0 \, \& \eta$, Ref.\refcite{GolLiu_proc}. The logic followed the observation that transverse virtual photons were significant, while the chiral even contributions, severely limited by the C-parity odd $t-$channel, were small, in principle. Once the dominant behaviour is established as due to transversity ($H_T(X,\zeta,t)$), the admixture of gluons at leading twist is forbidden. 
So focusing on charmed pseudoscalars provides access to purely intrinsic charm rather than photon-gluon fusion.

\section{Conclusions}
The emergence of the EIC as a top priority of the
nuclear physics community presents a unique opportunity to explore
the  non-perturbative sea and gluonic structure of the nucleon and nuclei. 
Here we presented a connection between intrinsic and open charm and 
transverse spatial distributions  of the nucleon's sea structure through 
electroproduction and GPDs.  

In particular the issue of  whether  perturbative gluons produce $c$ $\bar{c}$ pairs, or intrinsic, non-perturbative charm mechanisms are present in the nucleon, still remains unsettled.  
One possibility that we studied is that various asymmetries for exclusive charm production providing {\it e.g.} the azimuthal dependence for the $D$ or $D^*$ meson, the $cos(2\phi)$ term in the unpolarized differential cross section, may provide that discrimination. 

The proposed EIC presents a unique opportunity to test the distinction between 
intrinsic charm and competing models by exploiting its kinematic reach in $x$ and the
final state variables ($z$, and $P_T$)
in semi-inclusive DIS, and $t$ in exclusive proccesses.
The various design scenarios  range from $E_e/E_P=10/100-250\ {\rm GeV}$, $E_e/E_P=3-7/30-150 {\rm GeV}$ with luminosity projections
ranging from  
$\textrm{few} \times 10^{33} \, \textrm{cm}^{-2} \textrm{s}^{-1}$ 
to $\textrm{few} \times 10^{35} \, \textrm{cm}^{-2} \textrm{s}^{-1}$, 
and to a recently proposed medium energy
collider $E_e/E_P=3-7/30-60 \ {\rm GeV}$ (Ref.\refcite{Horn:2009cu}).

Another feature of exclusive charm production is that the produced charmed hyperon can carry significant polarization through a mechanism similar to strange hyperons. The $\Lambda_c$ 
hyperon's polarization 
is determined by its weak decay products and is known to provide favorable polarization analysis
(Ref.\refcite{DharGol}).
 Utilizing  polarized electron and 
proton beams in the proposed kinematic regime would allow the disentangling of the multiple mechanisms as well. We are studying these multiple spin correlation observables to see
what their impact can be in both LHC and EIC kinematical regimes (Ref.\refcite{GolLiu}).

\section*{Acknowledgements}
The authors appreciate the work of the organizers of Exclusives 2010. We also thank our colleague L. Gamberg for initial participation in this project.This work is partially supported by the U.S. Department
of Energy grants DE-FG02-01ER4120 (S.L.), and DE-FG02-92ER40702  (G.R.G.).


\begin{thebibliography}{9}

\bibitem{Nad} P.~M.~Nadolsky {\it et al.},
  Phys.\ Rev.\  D {\bf 78}, 013004 (2008)
 
\bibitem{Pum} J.~Pumplin, H.~L.~Lai and W.~K.~Tung,
  Phys.\ Rev.\  D {\bf 75}, 054029 (2007)
  
\bibitem{Collins} J.C. Collins, L. Frankfurt and M. Strikman, 
 Phys.\ Rev.\  D {\bf 56}, 2982 (1997)

\bibitem{AGL} S.~Ahmad, G.~R.~Goldstein and S.~Liuti,
  Phys.\ Rev.D{\bf 79}, 054014 (2009)
  
\bibitem{Bar}  V.~Barone, T.~Calarco and A.~Drago,
  Phys.\ Lett.\  B {\bf 390}, 287 (1997)

\bibitem{Vog}  W.~Vogelsang,
  Phys.\ Rev.\  D {\bf 57}, 1886 (1998)

\bibitem{Kum} M.~Hirai, S.~Kumano and M.~Miyama,
  Comput.\ Phys.\ Commun.\  {\bf 111}, 150 (1998); S.~Kumano and M.~Miyama;  
  Phys.\ Rev.\  D {\bf 56}, 2504 (1997)
  
\bibitem{Ana} L.~N.~Ananikyan and N.~Y.~Ivanov,
  Nucl.\ Phys.\  B {\bf 762}, 256 (2007)
 
\bibitem{Song} X.~Song,
  arXiv:hep-ph/9910515; {\it ibid} Int.\ J.\ Mod.\ Phys.\  A {\bf 18}, 1501 (2003).
 
\bibitem{AHLT1} S.~Ahmad, H.~Honkanen, S.~Liuti and S.K.Taneja,
  Phys.\ Rev.D{\bf 75}, 094003 (2007)

\bibitem{AHLT2} S.~Ahmad, H.~Honkanen, S.~Liuti and S.K.Taneja,
  Eur.\ Phys.\ J.\  C {\bf 63}, 407 (2009)

\bibitem{BelRad} A.~V.~Belitsky and A.~V.~Radyushkin,
  Phys.\ Rept.\  {\bf 418}, 1 (2005)
  
\bibitem{lat} G.~Bali, S.~Collins and A.~Schafer,
  PoS {\bf LATTICE2008}, 161 (2008)
 
\bibitem{Pate} S.~F.~Pate, D.~W.~McKee and V.~Papavassiliou,
  Phys.\ Rev.\  C {\bf 78}, 015207 (2008)
  
\bibitem{GolLiu_proc}  G.R. Goldstein, and S. Liuti, {\it these proceedings}.
\bibitem{Horn:2009cu}
  T.~Horn, P.~Nadel-Turonski and C.~Weiss,
  arXiv:0908.1999 [hep-ph].  
  
\bibitem{DharGol} W.~G.~D.~Dharmaratna and G.~R.~Goldstein,
  Phys.\ Rev.\  D {\bf 53}, 1073 (1996); G.~R.~Goldstein, arXiv:9907573 [hep-ph].
    
\bibitem{GolLiu} G.R. Goldstein, and S. Liuti, {\it in preparation}. 

\end{thebibliography}
\end{document}